\documentclass[aps,prl,twocolumn,showpacs,preprintnumbers,amsmath,amssymb]{revtex4}

\newcommand{\be}{\begin{equation}}
\newcommand{\ee}{\end{equation}}
\newcommand{\bea}{\begin{eqnarray}}
\newcommand{\eea}{\end{eqnarray}}
\newcommand{\rmd}{{\rm d}}

\usepackage{graphicx}
\usepackage{dcolumn}
\usepackage{bm}
\usepackage{color}

\newcommand{\e}{\xi}

\newcommand{\s}{\sigma}

\newcommand{\mycomma}{, }

\begin{document}

\title{Non-equilibrium frequency-dependent noise through a quantum dot: 
A real time functional renormalization group approach} 

\author{ C.P. Moca,$^1$ P. Simon,$^2$  C.H. Chung,$^3$ and G. Zar\'and,$^1$ }
\address{
$^1$Institute of Physics\mycomma Budapest University of Technology and Economics\mycomma  H-1521
Budapest\mycomma  Hungary\\
$^2$Laboratoire de Physique des Solides\mycomma  Univ. Paris Sud\mycomma  CNRS\mycomma  UMR 8502\mycomma  F-91405 Orsay Cedex\mycomma  France\\
$^3$Electrophysics Department\mycomma National Chiao-Tung University\mycomma
 HsinChu\mycomma Taiwan\mycomma R.O.C.\\
}
\date{\today}

\begin{abstract}
We construct a  real time current-conserving 
functional renormalization group (RG) scheme on the Keldysh contour 
to study frequency-dependent 
transport and noise through a quantum dot in the local moment regime.
We find that the current vertex  develops a non-trivial 
non-local structure in time, governed by a new set of RG equations. 
Solving these RG equations, we compute the complete 
frequency and temperature-dependence of the noise spectrum. 
For  voltages large compared to  the Kondo temperature, $eV\gg k_BT_K$, two sharp anti-resonances  
are found in the noise spectrum at frequencies $\hbar \omega = \pm e V$,  and correspondingly, 
two peaks in the ac conductance through the dot. 
\end{abstract}

\maketitle

{\emph{ Introduction.}}
Not only the current flowing through a system, but also  
its fluctuations (noise) carry crucial information on the physics that
governs transport~\cite{landauer}. Zero-frequency noise (shot noise) has been
used, e.g., to reveal the fractional charge  of the quasiparticle
excitations in the fractional quantum Hall liquid~\cite{glattli}. 
The low-frequency electrical noise has been extensively studied
in various systems\cite{blanter} and is by now relatively well
understood.  However, even more information is stored in the
finite-frequency (FF) current noise:  It has been predicted that the
FF noise is sensitive to the statistics of the quasiparticles
\cite{statistics}  and a crossover between different quantum
statistics can be
potentially observed as function of frequency, 
similar to the one observed as a function of temperature~\cite{Heiblum}.
Moreover, in the quantum regime characterized by  frequencies higher
than the applied voltage or temperature,  
the FF noise is a powerful tool to reveal the characteristic time scales of the probed 
system \cite{gabelli} as well as the dynamics of the excitations or
the importance of interactions.

Due to their small size, transport through quantum dots (QD) is strongly
influenced  by the Coulomb blockade.  
In particular, QDs with an odd number of electrons behave as artificial magnetic
impurities and  exhibit the Kondo effect \cite{david}, a paradigmatic many-body
phenomenon corresponding to the screening  
of the spin of the quantum dot  by the conduction electrons of the
leads at temperatures $T$ below the Kondo temperature, $T_K$. 
QDs thus provide an ideal test ground to study non-equilibrium
transport in the presence of strong interactions. 
While the conductance  of a QD in the Kondo
regime is well understood by now
\cite{glazman-review},  much less is known about
current-fluctuations. Though these are promising quantities
to characterize   the out of equilibrium Kondo effect,
nevertheless, most experiments focused so far on measurements of the
average current 
 \cite{franceschi02,paaske06,grobis08}, and even results on {\it
   low-frequency} noise measurements  
have only appeared recently \cite{delattre09}. While the effect of ac
voltage on the non-equilibrium Kondo effect has been studied
experimentally relatively longtime ago~\cite{kogan04},
no FF noise measurements have been
reported so far to our knowledge. Theoretically, most studies
focused on shot noise: a 
non-monotonous bias-dependence of the shot noise 
 with a maximum at $eV\sim k_BT_K$ has been found at $T\ll T_K$ \cite{meir02},  
and a universal  ratio  $5/3e$ between the shot noise and  the backscattering current at $T=0$ 
has been predicted  for the $SU(2)$ Kondo effect \cite{sela06}.

The purpose of this work is to provide a general analysis of the
finite-frequency current noise through a
quantum dot in the local moment regime. To achieve this, we construct a  real time
functional renormalization group (FRG) scheme on the Keldysh contour 
to study frequency-dependent 
transport and noise through a Kondo quantum dot. Our formalism reproduces 
 the scaling equations of Rosch et al. \cite{rosch} 
for the vertex function. However, we find that
the current vertex also develops a non-trivial 
{\it non-local} structure in time, governed by a new set of RG equations.
Such structure of the current vertex turns out to be unavoidable  to guarantee current
conservations and is necessary to calculate the finite frequency
current noise  in a controlled manner. Solving this set of RG
equations, we compute the complete frequency and temperature dependent 
noise spectrum through the dot. Our approach is valid at any
frequency $\omega$,  voltage $V$ and temperature $T$ provided
that ${\rm max}\{eV,k_BT\}> k_B T_K$.   
For  frequencies $\hbar \omega\gg k_B T_K$, we find  sharp anti-resonances  
in the voltage dependence of the noise spectrum at $eV =
\hbar \omega$,
which  gradually disappear with increasing
temperature. The absorption noise is also found to exhibit strong
anomalies at $\hbar \omega =  eV$, and we find similar
anomalies  in the non-equilibrium ac conductance, too, where a 
split non-equilibrium Kondo resonance is observed. 
Precursors of the noise anomaly have been found
in the zero-temperature symmetrized noise at finite frequency, 
as first computed at the Toulouse point of the Kondo
model~\cite{schiller}, and later confirmed 
by a non-equilibrium one-loop perturbative
calculation~\cite{schoeller}. However, logarithmic singularities are
completely absent at the rather special Toulouse
point~\cite{schiller},
 while the method of Ref.~\cite{schoeller} was not accurate enough to
 capture fine details  of the anomaly.

{\emph{ Model.}} In this paper, we focus our attention to the local moment 
regime of the quantum dot, where we can describe the electron 
on the dot as a spin $S=1/2$ moment which couples to electrons in the left 
and right electrodes through an exchange interaction~\cite{glazman-review},
\begin{equation}
H_{\rm int}=  \frac 1 2 \sum_{\alpha,\beta=L,R} \sum_{\sigma,\sigma'}
j_{\alpha\beta}\;
\mathbf{S} \;\psi_{\alpha\sigma}^\dagger {\mathbf{\mathbf \sigma}}_{\sigma\sigma'}
\psi_{\beta\sigma'}\;.
\label{eq:H_int}
\end{equation}
Here $\mathbf \sigma$ stands for the three Pauli matrices, 
the fields $\psi_{\alpha\sigma} = \int  c_{\alpha\s}(\e) e^{-|\e| a}\;d\e$
destroy electrons of spin $\sigma$ in leads $\alpha\in\{L,R\}$, 
with $1/a$ a high energy cut-off~\footnote{The annihilation operators
  $c_{\alpha\s}(\e)$ satisfy  
 $\{c^{\dagger}_{\alpha\s}(\e),c_{\alpha'\s'}(\e')\}
=\delta_{\alpha\alpha'}\delta_{\s\s'}\delta(\e-\e')$.} .
The dynamics of  $\psi_{\alpha\sigma}$
is governed by the non-interacting Hamiltonian,
$H_0 = \sum_{\alpha,\s} \int {\rm d}\e\; (\e + \mu_\alpha)\; 
c^{\dagger}_{\alpha\s}(\e) c_{\alpha\s}(\e) 
$, with $\mu_\alpha=eV_\alpha$, the chemical potential shift of 
lead $\alpha$.

 To describe the spin using standard field-theoretical methods,
we make use of Abrikosov's pseudo-fermion
representation~\cite{Abrikosov}: 
we introduce a fermion operator $f^\dagger_s$ for each spin 
component, $s=\pm 1/2$, and represent the spin operator  as
$\hat S^i\to \sum_{s,s'}\frac 1 2 \; f^\dagger_s \s^i_{s,s'}f_{s'}$
with the additional  constraint, $ \sum_{s}f^\dagger_s f_{s}=1$.

 We then employ a path integral 
formalism on the Keldysh contour. In this approach each fermionic field 
is replaced by two time-dependent Grassmann fields living 
on the upper and lower  Keldysh contour ($\kappa=1,2$), respectively, 
and the dynamics 
is determined by the Keldysh action, 
${\cal S} = {\cal S}_{\rm lead} +{\cal S}_{\rm spin} +{\cal S}_{\rm int}$. 
The parts ${\cal S}_{\rm lead}$ and ${\cal S}_{\rm spin}$ describe the 
conduction electrons and the spin in the absence of interaction. They
are quadratic in the fields, and 
determine the non-interacting Green's functions~\cite{epaps}.

The interaction part of the action, ${\cal S}_{\rm int}$, is diagonal  in the Keldysh 
indices and is initially local in time. However,
elimination of high energy degrees of freedom in course of the 
 RG procedure generates retardation effects, 
and the interaction becomes {\em non-local}. We find that, 
 with a good approximation, it  can be expressed 
as
\bea 
{\cal S}_{\rm int} &=& \sum_\kappa  \sum_{\alpha,\beta}
s_\kappa\;\frac 14  \int {\rm d}t_1\;{\rm d}t_2\;g_{\alpha\beta}(t_1-t_2)\;
\nonumber
\\
&& \bar f^\kappa(T_{12}) \vec \sigma f^\kappa(T_{12})  \cdot
\bar \psi^\kappa_\alpha(t_1) \vec \sigma \psi_\beta^\kappa(t_2) \;, 
\label{eq:S_int}
\eea
where $T_{12}= (t_1 + t_2)/2$ and $s_\kappa = \pm 1$ for the upper and 
lower Keldysh contours, respectively. The initial (bare)  
coupling function $g_{\alpha\beta}(t)$ is local in time, and is given by
$g^{(0)}_{\alpha\beta}(t)
= j_{\alpha\beta}\;\delta(t)$. 
The justification for this  
structure,  Eq.~\eqref{eq:S_int}, is straightforward: The spin evolves 
very slowly, and its time evolution 
can be very well approximated by the one in the absence of 
interactions at electronic time scales. However, conduction electrons 
have fast dynamics, and their retardation effects become important as one 
approaches smaller energy scales.

{\emph{ Functional RG.}}
We construct the RG equations by expanding the action in 
${\cal S}_{\rm int}$ and rescaling the cutoff parameter $a\to a'$. 
An integro-differential equation is obtained 
for the functions $g_{\alpha\beta}(t)$,
 which becomes simple in Fourier space, 
\be 
\frac {\rmd  {\bf g} (\omega)}{\rmd l}=
{\bf g} (\omega)\; {\bf q}(\omega,a)\;{\bf g} (\omega)\;.
\label{eq:scaling_g}
\ee
Here $l = \ln(a/a_0)$ is the scaling variable, $a_0$ is the initial value 
of the cut-off time, and we introduced the matrix notation, 
$g_{\alpha\beta}\to {\bf g}$. The matrix $ {\bf q} (\omega,a)$
is a cut-off function, which depends somewhat on the precise cut-off 
scheme, but for practical purposes is well-approximated by the function
$q_{\alpha\beta}(\omega,a)\approx \delta_{\alpha\beta}\; \Theta(\frac 1 a - 
|\omega-\mu_\alpha|)$ at $T=0$ temperature~\cite{rosch}. 
The scaling equation, Eq.~\eqref{eq:scaling_g}, is identical to the one 
obtained in a more heuristic  way in Ref.~\cite{rosch}, however, in our 
real time functional RG formalism the derivation is rather straightforward and simple~\cite{long}. 
We remark that the usual poor man's RG
procedure can be  recovered by dropping the 
time-dependence of $g_{\alpha\beta}$, and replacing the generated 
non-local couplings 
by local ones,  $g_{\alpha\beta}(t) \to  \delta(t)\; \int \rmd t\; 
g_{\alpha\beta}(t)$, which corresponds to setting $\omega\to0$
 in Eq.~\eqref{eq:scaling_g}.

Our primary purpose is to compute current-current correlation 
functions. To do that, we first define the left and right 
current operators from the equation of motion, giving 
$\hat I_L (t) = - \hat I_R (t)  = \sum_{\alpha\beta}\frac e 
2 v^L_{\alpha\beta}\hat {\bf S}(t) \cdot\hat \psi^\dagger_{\alpha}(t) 
\mathbf{\sigma} \hat \psi_{\beta}(t)$, with the current vertex
matrices  defined as
\be 
{\bf v}^L = -{\bf v}^R=
\begin{pmatrix}
0 & - i \; j_{LR} \\
i \; j_{LR} & 0
\end{pmatrix}
\;.
\ee

In the path integral language, it is useful to introduce a corresponding generating functional, 
\be 
Z[h_\alpha^\kappa(t)] \equiv 
\langle e^{-i \;\sum_{\kappa,\alpha} \int \rmd t \;  h_\alpha^\kappa(t) I^\kappa_\alpha (t)}\rangle_{\cal S}\;,
\label{generating_functional}
\ee
from which the current-current correlation functions can be generated by functional differentiation with respect to $h_\alpha^\kappa(t)$.
A systematic investigation of the 
leading logarithmic  diagram series shows that the expression 
of the current field,  $I^\kappa (t)$,  necessarily becomes 
\emph{non-local in time} under the RG procedure, and acquires a form,
\bea 
I^\kappa_L(t) &=& \frac {e} 4\sum_\kappa  \sum_{\alpha\beta}
\int {\rm d}t_1{\rm d}t_2\;V_{\alpha\beta}^L(t_1-t,t-t_2,a)\;
\nonumber
\\
&& \bar f^\kappa(t) \vec \sigma f^\kappa(t)  \cdot
\bar \psi^\kappa_\alpha(t_1) \vec \sigma \psi^\kappa_\beta(t_2) \;. 
\eea
%
The physical motivation of the double time-structure is 
simple: in the renormalized theory it is not enough to 
know the times electrons  enter and leave the dot
($t_{1,2}$), but the time  
of the current measurement, $t$, must also be  kept track of.

It is relatively straightforward to derive the 
scaling equations from the perturbative expansion of 
Eq.~\eqref{generating_functional}, and we obtain
\bea 
\frac {\rmd  {\bf V}^L (\omega_1,\omega_2)}{\rmd l} &=&
 {{\bf V}^L (\omega_1,\omega_2)}\; {\bf q}(\omega_2,a)\;{\bf g} (\omega_2)\;
\nonumber \\ 
&+&
{\bf g} (\omega_1)\; {\bf q}(\omega_1,a)\; 
{{\bf V}^L (\omega_1,\omega_2)}\;.
\label{eq:scaling_v}
\eea 
This equation needs be solved parallel to the scaling equation, 
Eq.~\eqref{eq:scaling_g} with the boundary condition, 
${\bf V}^{L/R} (\tau_1,\tau_2,a_0) = \delta(\tau_1)\;\delta(\tau_2)
\;{\bf v}^{L/R}$. Though 
the renormalized couplings ${\bf g}_L(\omega)$ drive the scaling 
of the current vertexes, ${\bf V}_L (\omega_1,\omega_2,a_0)$, there seems
 to be no simple connection between these too. In other words, 
it is unavoidable to introduce the renormalized current vertexes within 
the  functional RG scheme to compute time-dependent 
current correlations. The above extension seems to be also necessary 
to guarantee {\em current conservation}:  Eq.~\eqref{eq:scaling_v} is  
linear in $ {\bf V}^L$, and therefore the condition 
$I^\kappa_L(t) + I^\kappa_R(t)\equiv 0$ is automatically satisfied,  
for any value of the cut-off, $a$. On the other hand, we could not 
find any way to generate a current field from just the renormalized 
action, Eq.~\eqref{eq:S_int}, such that it respects current 
conservation.

\begin{figure}
\includegraphics[width=0.75\columnwidth,clip]{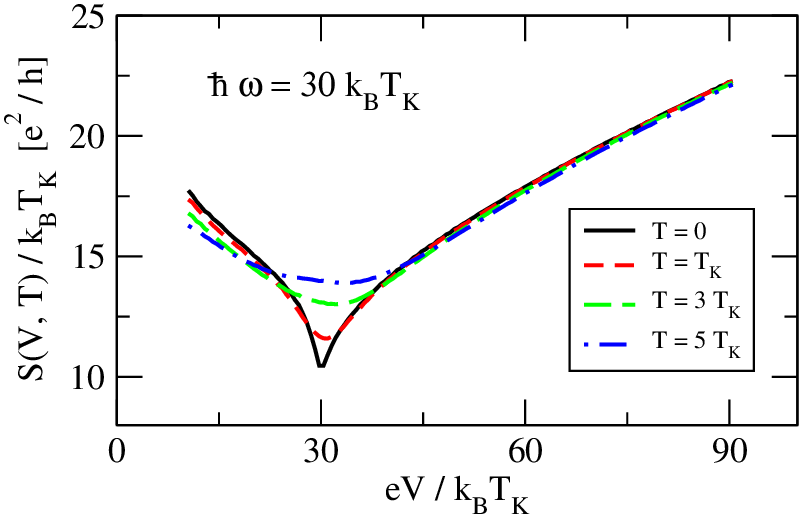}
\caption{\label{fig:noise_Vdep} (color online)
Voltage-dependence of the symmetrical noise,
 as computed by FRG for $\hbar\omega=30\; k_B T_K$. }
\end{figure}
We solved  Eqs.~\eqref{eq:scaling_g} and
\eqref{eq:scaling_v} numerically to obtain   ${\bf g}$ and ${\bf
  V}^{L/R}$. Both display singularities at 
frequencies $\hbar \omega=\pm eV/2$ \cite{epaps}.
With the couplings  ${\bf g} (\omega_1)$ and the current vertexes 
${\bf V}^{L/R} (\omega_1,\omega_2)$ in hand, we then proceeded to compute the 
noise through the device, by doing perturbation theory with the renormalized 
action. For the Fourier transform of the absorption and emission 
noise components $S_{LL}^{>}(t) \equiv \langle \hat I_L(t)\hat I_L(0)\rangle$, 
and $S_{LL}^{<}(t) \equiv \langle \hat I_L(0)\hat I_L(t)\rangle$,
we obtain $S_{LL}^{>}(\omega) =  S_{LL}^{<}(-\omega) $, with 
\bea 
S_{LL}^{>}(\omega) &=& \frac{e^2} 2\;
S(S+1)\int \frac{\rmd \tilde \omega}{2\pi}\;{\rm Tr}\{
{\bf V}^L(\tilde\omega_-,\tilde\omega_+)
\nonumber
\\
&& {\bf G}^{>}(\tilde\omega_+)
{\bf V}^L(\tilde\omega_+,\tilde\omega_-)  {\bf G}^{<}(\tilde\omega_-)
\}\;.
\eea
Here $\tilde\omega_\pm = \tilde \omega \pm \frac \omega 2$, and 
the bigger and lesser Green's functions 
$ G_{\alpha\beta}^{>/<}(\omega) = \pm i\;2\pi\;
\delta_{\alpha\beta}\;f(\pm(\omega-\mu_\alpha))$.

{\emph{Results.}}
The symmetrized noise spectrum, $S_{LL}(\omega) \equiv 
\frac 1 2 [S^>_{LL}(\omega) + S^<_{LL}(\omega)]$ is plotted 
in Fig.~\ref{fig:noise_Vdep} for $\hbar \omega= 30 \;k_B T_K$ as a
function of voltage, $V$.  Clearly, the noise spectrum 
shows rather strong features  at the bias voltages, $eV\approx \hbar
\omega$.  The appearing dips are clear fingerprints of the non-equilibrium Kondo
effect, and they gradually vanish as we increase the temperature, $T$  
\footnote{Finite temperature has been included through the spin
  relaxation time, similar to Refs.~\cite{rosch}.}.
The  increase in the noise with decreasing $V$ at low voltages 
can be understood as being
the consequence of increasing spin relaxation time, resulting in 
an increased Kondo conductance through the dot. At high voltages, on the other
hand, increased photon emission is mainly responsible for the increasing
noise, thereby giving the V-shaped pattern. 

\begin{figure}
\includegraphics[width=0.75\columnwidth,clip]{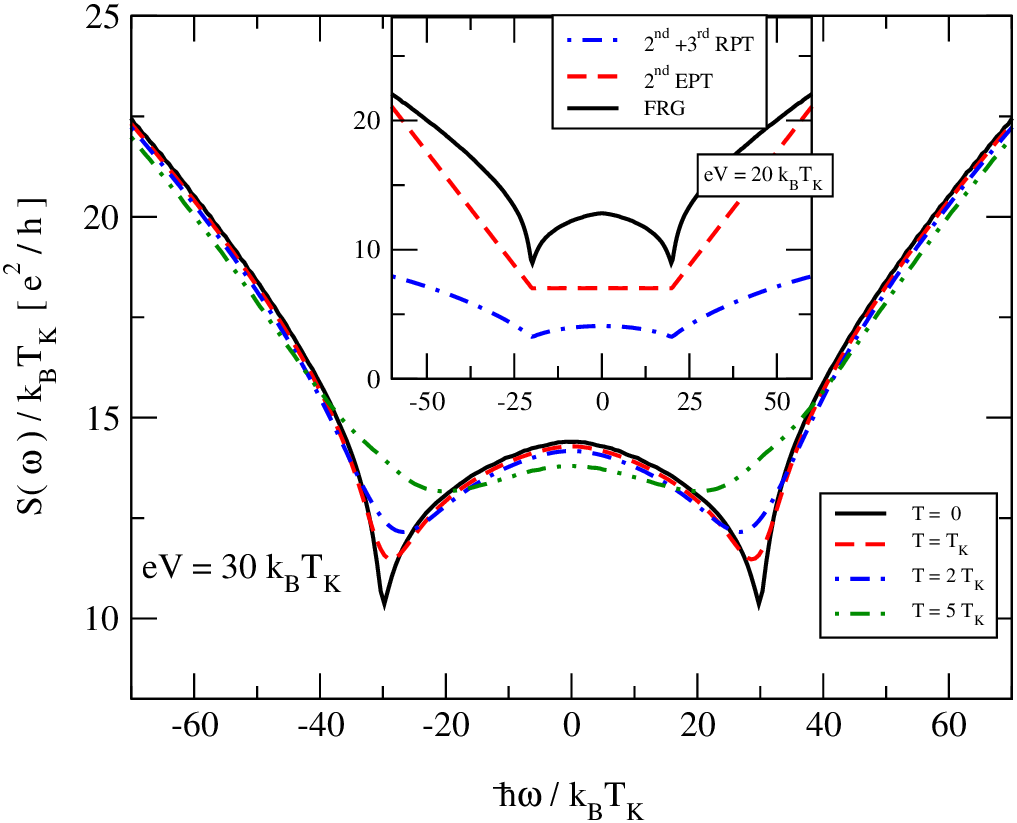}
\caption{\label{fig:noise} (color online)
Temperature-dependence of the symmetrical noise,
 as computed by FRG. Inset: Comparison with third 
order renormalized  perturbation  theory (RPT) using a reduced 
cut-off, $\tilde a = \hbar/10eV$ and the corresponding 
couplings  $j_{LR}(\tilde a)$, and effective second order 
perturbation theory with renormalized couplings 
$j_{LR}(\tilde a = \hbar/ eV)$, but still using the original 
bandwidth. None of these methods are able to get close to the 
FRG results. 
}
\end{figure}

Similar features appear in the
frequency-dependent symmetrized noise (see Fig.~\ref{fig:noise}).
It is instructive to compare the FRG results with
perturbation theory, giving 
\bea
S^>_{LL}(t)&=& - e^2\;\frac 3 4 |j_{LR}|^2 \cos(eVt)\Bigl\{ 
\frac 1 {(t - i\;a)^2}
\nonumber 
\\
&+& 2 (j_{LL}+ j_{RR}) \frac {\ln(1 + i \;t/a) } {t (t - 2 \;i\; a)}
+\dots \Bigr\}\;.
\eea
The curves in the inset of Fig.~\ref{fig:noise} are obtained
by taking the Fourier transform of this expression. 
While the perturbative result also exhibits singular features 
at $\hbar \omega = \pm eV$, however, it does not 
reproduce the precise shape of the anomaly, even if we use renormalized 
parameters, $j_{\alpha\beta}\to j_{\alpha\beta}(eV)$ obtained 
by solving the usual leading logarithmic scaling equations. 

\begin{figure}
\includegraphics[width=0.75\columnwidth,clip]{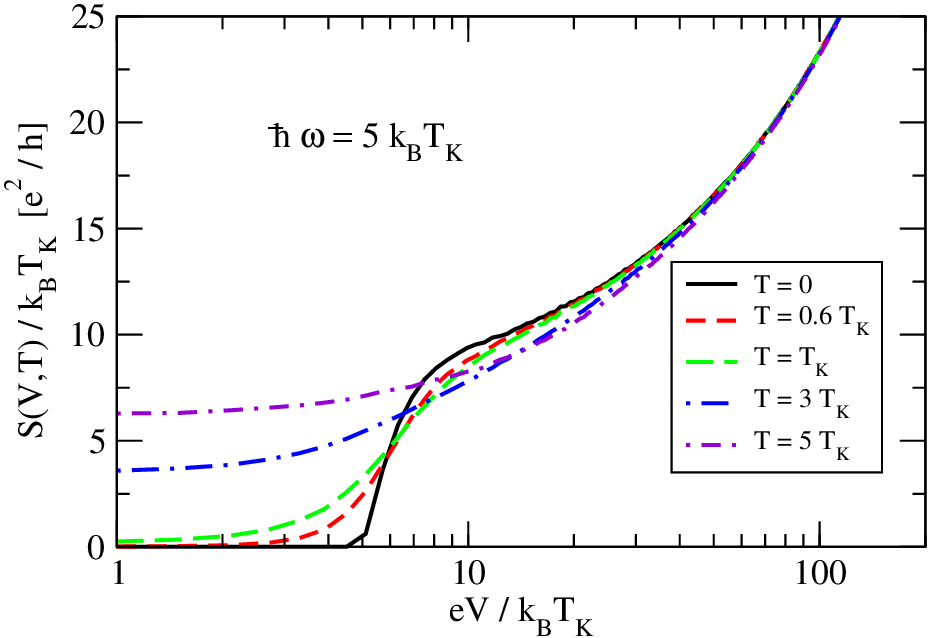} 
\caption{\label{fig:noise_emission} (color online)
Voltage and temperature-dependence of the emission noise 
as computed through FRG for $\hbar\omega=5 k_B T_K$. 
}
\end{figure}
Experimentally, it may be more convenient to measure
separately the emission 
or absorption noise components 
at a fixed finite frequency, $S_{e/a}(\omega>0)\equiv
S^{<}(\pm \omega)$, as function of the bias
voltage \cite{basset}.  
As shown in Fig.~\ref{fig:noise_emission},
at $T=0$ the emission noise vanishes at voltages $V<\hbar\omega/e$ 
due to energy conservation, and has an abrupt logarithmic singularity
at $V=\hbar\omega/e$ at temperatures $T\ll T_K$, which is gradually
smeared out for  $k_B T > eV$. 

Another quantity which may be easier to access experimentally is  the
non-equilibrium  finite frequency linear conductance, defined 
as the current response of the system to an external 
time-dependent variation of one of the lead potentials. 
According to a formula of Safi~\cite{safi}, this can be expressed as
\be 
{\rm Re}\, G_{LL}(\omega,V) = \frac 1 {\hbar \omega} 
(S_{LL}^>(\omega) - S_{LL}^<(\omega))
\;.
\ee 

\begin{figure}[b]
\includegraphics[width=0.75\columnwidth,clip]{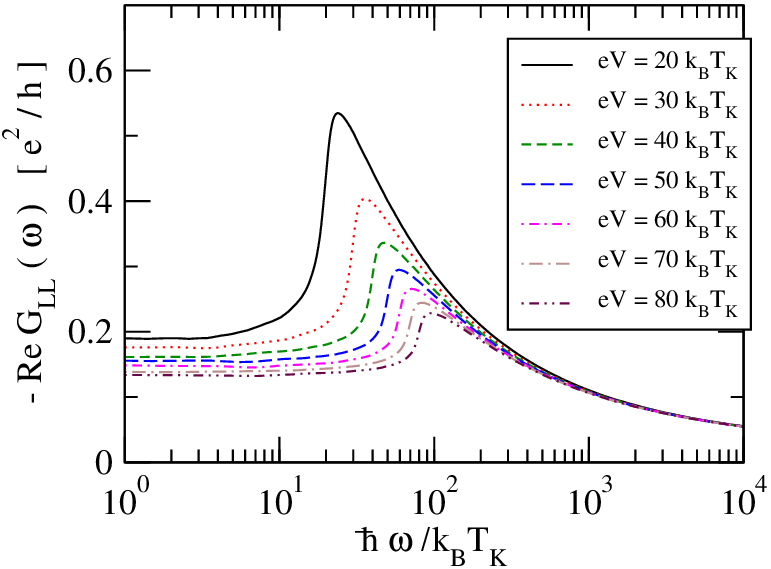}
\caption{\label{fig:ac_cond} 
(color online) The ac conductance 
for  $\omega > 0$.}
\end{figure}  

Notice that $G_{LL}(0,V)$ is just the usual non-equilibrium differential 
conductance, $G_{LL}(0,V)= \partial_{V_L}\langle\hat I_L\rangle$,
while $G_{LL}(\omega, V\to 0)$ corresponds to the usual 
equilibrium ac conductance~\cite{sindel}.
${\rm Re}\, G_{LL}(\omega,V)$ is an even function of 
$\omega$ and exhibits two  peaks 
at $\hbar \omega = \pm e V$, associated with the non-equilibrium 
Kondo effect~\cite{meir02} (see Fig.~\ref{fig:ac_cond}) . This confirms that
 the dips in the noise are related to the splitting of the Kondo
 resonance at a finite bias.

{\emph{ Summary.}} 
We have developed a real-time functional renormalization
group approach.  We have shown that the current vertex becomes
non-local in time under renormalization which turns out to be
necessary in order to ensure non-equilibrium current
conservation. Within this formalism, we have been able to  
calculate  the voltage and temperature dependence of the current noise
at finite frequency and the   non-equilibrium ac conductance,
quantities which are within experimental reach~\cite{basset}.

{\emph{Acknowledgment.}}
We would like to thank 
J. Basset, R. Deblock, H. Bouchiat, and B. Reulet
for interesting discussions. 
This research has been supported by Hungarian grants OTKA Nos.
NF061726, K73361, Romanian grant CNCSIS PN II ID-672/2008, the 
EU  GEOMDISS project, and Taiwan's NSC grant No.98-2918-I-009-06, 
No.98-2112-M-009-010-MY3, the MOE-ATU program and the NCTS 
of Taiwan, R.O.C.. 
C.H.C. acknowledges the hospitality of Yale University. 



\end{document}